\input harvmac
% -------------------------AK Definitions
\font\cmss=cmss10 \font\cmsss=cmss10 at 7pt
 \def\inbar{\,\vrule height1.5ex width.4pt depth0pt}
\def\IZ{\relax\ifmmode\mathchoice
{\hbox{\cmss Z\kern-.4em Z}}{\hbox{\cmss Z\kern-.4em Z}}
{\lower.9pt\hbox{\cmsss Z\kern-.4em Z}}
{\lower1.2pt\hbox{\cmsss Z\kern-.4em Z}}\else{\cmss Z\kern-.4em
Z}\fi}
\def\IB{\relax{\rm I\kern-.18em B}}
\def\IC{{\relax\hbox{$\inbar\kern-.3em{\rm C}$}}}
\def\ID{\relax{\rm I\kern-.18em D}}
\def\IE{\relax{\rm I\kern-.18em E}}
\def\IF{\relax{\rm I\kern-.18em F}}
\def\IG{\relax\hbox{$\inbar\kern-.3em{\rm G}$}}
\def\IGa{\relax\hbox{${\rm I}\kern-.18em\Gamma$}}
\def\IH{\relax{\rm I\kern-.18em H}}
\def\II{\relax{\rm I\kern-.18em I}}
\def\IK{\relax{\rm I\kern-.18em K}}
\def\IP{\relax{\rm I\kern-.18em P}}
\def\IR{\relax{\rm I\kern-.18em R}}

\def\tilde{\widetilde}
\noblackbox
%----------------------------

%%Some math defs
\def\edkps{E_{{\rm DKPS}}}
%-------------------------

\Title{\vbox{\baselineskip12pt\hbox{hep-th/9712223} 
\hbox{HUTP-97/A053, LBNL-41190} 
\hbox{SLAC-PUB-7724, UCB-PTH-97/67}
}}
{\vbox{\centerline{
On the Matrix Description of}
\smallskip
\centerline{Calabi-Yau Compactifications}
}}
\centerline{Shamit Kachru$^{1}$, Albion Lawrence$^{2}$
and Eva Silverstein$^{3}$}
\bigskip
\centerline{$^{1}$Department of Physics}
\centerline{University of California at Berkeley}
\smallskip
\centerline{and}
\smallskip
\centerline{Theoretical Physics Group}
\centerline{Lawrence Berkeley National Laboratory}
\centerline{University of California}
\centerline{Berkeley, CA 94720}
\bigskip
\centerline{$^{2}$ Department of Physics}
\centerline{Harvard University}
\centerline{Cambridge, MA 02138}
\bigskip
\centerline{$^3$ Stanford Linear Accelerator Center}
\centerline{Stanford University}
\centerline{Stanford, CA 94309}

\medskip
\bigskip
\noindent
We point out that the matrix description of
M-theory compactified on Calabi-Yau threefolds
is in many respects simpler than
the matrix description of a $T^6$ compactification.
This is largely because of the differences between 
D6 branes wrapped on Calabi-Yau threefolds and D6 branes
wrapped on six-tori.
In particular, if we define the matrix theory 
following the prescription of Sen
and Seiberg, we find that the remaining degrees of freedom are 
decoupled from gravity. 

\medskip
\Date{December 1997}
%\draftmode

\lref\othersix{A. Losev, G. Moore, and S. Shatashvili, \lq\lq M \& m's,"
hep-th/9707250\semi 
I. Brunner and A. Karch, \lq\lq Matrix Description of M-theory on 
$T^6$,'' hep-th/9707259\semi
A. Hanany and G. Lifschytz, \lq\lq M(atrix) Theory on $T^6$ and a 
m(atrix) Theory Description of KK Monopoles,'' hep-th/9708037.}
\lref\bfss{T. Banks, W. Fischler, S. Shenker, and L. Susskind, 
``M theory as a Matrix Model:  A Conjecture," Phys. Rev. {\bf D55}
(1997) 5112, hep-th/9610043.}
\lref\dkps{M. Douglas, D. Kabat, P. Pouliot, and S. Shenker, 
``D-branes and Short Distances in String Theory,'' Nucl. Phys.
{\bf B485} 85, 
hep-th/9608024.}
\lref\hellpol{S. Hellerman and J. Polchinski, \lq\lq Compactification
in the Lightlike Limit," hep-th/9711037.}
\lref\sen{A. Sen, ``D0-branes on $T^n$ and Matrix Theory,'' 
hep-th/9709220.}
\lref\seiberg{N. Seiberg, ``Why is the Matrix Model Correct?,''
Phys. Rev. Lett {\bf 79} 3577, 
hep-th/9710009.}
\lref\cdgp{P. Candelas, X. De la Ossa, P. Green, and L. Parkes,
``A Pair of Calabi-Yau Manifolds as an Exactly Soluble Superconformal
Theory,'' Nucl. Phys. {\bf B359} (1991) 21.}
\lref\gk{B. Greene and Y. Kanter, ``Small Volumes in Compactified String
Theory", Nucl. Phys. {\bf B497} 127, hep-th/9612181.} 
\lref\andy{A. Strominger, ``Massless Black Holes and Conifolds in
String Theory,'' Nucl. Phys. {\bf B451} 109, hep-th/9504090.}
\lref\lenny{L. Susskind, ``Another Conjecture about Matrix Theory,''
hep-th/9704080.}
\lref\wati{W. Taylor IV, \lq\lq D-brane Field Theory on Compact
Space,'' Phys. Lett. {\bf B394} (1997) 283, hep-th/9611042.}
\lref\cydist{P. Aspinwall, B. Greene, and D. Morrison, ``Measuring
Small Distances in N=2 Sigma Models,'' Nucl. Phys. {\bf B420} (1994)
184, hep-th/9311042\semi
P. Aspinwall, ``Minimum Distances in Nontrivial String Target
Spaces,'' Nucl. Phys. {\bf B431} (1994) 78, hep-th/9404060.} 
\lref\syz{A. Strominger, S. Yau, and E. Zaslow, ``Mirror Symmetry
is T-duality,'' Nucl. Phys. {\bf B479} (1996) 243, hep-th/9606040.}
\lref\edphases{E. Witten, ``Phases of N=2 Theories in Two Dimensions,"
Nucl. Phys. {\bf B403} (1993) 159, hep-th/9301042.} 
\lref\silvwitt{E. Silverstein and E. Witten, ``Criteria for Conformal
Invariance of (0,2) Models,'' Nucl. Phys. {\bf B444} 
(1995) 161, hep-th/9503212.}
\lref\albionetal{T. Jayaraman, A. Lawrence, H. Ooguri and
T. Sarkar, to appear.}
\lref\seibfive{N. Seiberg, \lq\lq New Theories in Six Dimensions 
and Matrix Description
of M-theory on $T^5$ and $T^5/Z_2$," hep-th/9705221.}
\lref\dvvfive{R. Dijkgraaf, E. Verlinde and H. Verlinde, \lq\lq BPS Spectrum
of the Five-brane and Black Hole Entropy," Nucl. Phys. {\bf B486} 77,
hep-th/9603126; and "BPS Quantization of the Five-brane," Nucl.
Phys. {\bf B486} 89, hep-th/9604055.}
\lref\elitzuretal{S. Elitzur, A. Giveon, D. Kutasov and E. Rabinovici,
\lq\lq Algebraic aspects of Matrix Theory on $T^d$," hep-th/9707217}
\lref\andyjoe{J. Polchinski and A. Strominger, \lq\lq New Vacua for Type
II String Theory," Phys. Lett. {\bf B388} 736.}
\lref\bsv{M. Bershadsky, V. Sadov and C. Vafa, 
\lq\lq D-strings on D-manifolds,"
Nucl. Phys. {\bf B463} 398, hep-th/9510225.}
\lref\horioz{K. Hori and Y. Oz, \lq\lq F-Theory, T-duality on
K3 Surfaces and N=2 Supersymmetric Gauge Theories in Four Dimensions,"
Nucl. Phys. {\bf B501} 97, hep-th/9702173.}
\lref\gepneropen{A. Recknagel and V. Schomerus, 
\lq\lq D-branes in Gepner Models,"
hep-th/9712186.}
\lref\do{M. R. Douglas and H. Ooguri, \lq\lq Why Matrix Theory is Hard,"
hep-th/9710178.}
\lref\dos{M.R. Douglas, H. Ooguri and S.H. Shenker, 
\lq\lq Issues
in M(atrix) Theory 
Compactification," Phys. Lett. {\bf B402} 36, hep-th/9702203.}

\newsec{Introduction}

The matrix model proposal for M theory \bfss\ aims to describe a space-time
theory with gravity via a much simpler theory which is manifestly better-defined.
With no dimensions compactified, the theory
is described by the large-N quantum mechanics of
D0-branes in the limit $g_s \to 0$ (vanishing string coupling).
However, it has become clear that compactification
requires the inclusion of additional degrees of
freedom.  Thus, to learn what degrees of freedom are needed in the
full formulation of the theory, we require
an understanding of the fully compactified theory (or theories, depending
on the level of connectedness of M theory backgrounds).  
The degrees of freedom that are needed in various other simple
compactifications 
have been determined.  On tori of dimension $p\le 4$ the matrix description
can be written as a quantum field theory (for $p\le 3$ it is
$p+1$-dimensional Yang-Mills theory with
16 supercharges \refs{\bfss,\wati}).  On $T^5$ the theory
can be defined as the \lq\lq little string theory" on a wrapped
fivebrane \refs{\seibfive,\dvvfive}.

The M-theory  compactification on
$T^6$, however, seems overly complicated \refs{\sen,\seiberg}.
One finds that the finite-N DLCQ description of M theory on 
$T^6$ is given by a theory of
ALE sixbranes (that is, of the $6+1$-dimensional
core of an $A_{N-1}$ ALE singularity in
eleven dimensions) in M-theory \refs{\othersix,\sen,\seiberg}, which 
fails to decouple from bulk gravity \refs{\sen,\seiberg}.  We
are then left with the unappetizing prospect of defining the
6-dimensional compactification of M-theory via M-theory
({\it cf} also \elitzuretal ).

One of the promising features of the matrix model is the independence
of its motivating arguments from supersymmetry.
Both the infinite momentum frame and DLCQ arguments for
decoupling of $p_{11}\le 0$ modes do not refer to supersymmetry.
In particular, the prescription of \refs{\lenny,\seiberg} 
applies to any compactification.  However, one might
expect that breaking supersymmetry complicates
matrix theory compactifications even further.

Nonetheless, we show in this note that for compactifications on
six-manifolds, the situation in fact simplifies
upon reducing the amount of supersymmetry.  In particular, the
difficulties with $T^6$
compactifications do not persist for generic six-manifold compactifications with
eight space-time supercharges (for which the D0-brane 
theory has four supercharges).
Using well-known facts about
Calabi-Yau compactifications, 
we find that the DLCQ compactification of M-theory on
a generic Calabi-Yau threefold following \refs{\lenny,\sen,\seiberg}
is described by a theory decoupled from gravity.  When we need a concrete
example we will always use the quintic threefold in $\IP^4$, but
we expect our results to 
generalize in a straightforward way to many other
Calabi-Yau spaces.\foot{In the special case of toroidal
orbifolds and their deformations, the analysis will be somewhat different.}
Our discussion relates issues of singularity resolution, which have played
a large role in the string duality story, to matrix theory.

\newsec{$T^6$ vs $CY_3$}

\subsec{Review of the $T^6$ case}

Let us review first the situation on $T^6$ \refs{\sen,\seiberg}.    
One considers N D0-branes in type IIA theory on a six-torus 
whose cycles are each of fixed size in
eleven-dimensional Planck units, in the
limit of vanishing string coupling ($g_s \to 0$).
In this limit, we wish to focus on states with
energy of the order $g_s^{2/3}/ \ell_P$, where $\ell_P$ 
is the eleven-dimensional Planck scale; these correspond to
M-theory states with finite light-cone energy.
We are therefore in the DKPS regime \dkps,
and we will call this energy $\edkps$.
In the four noncompact dimensions, the D0-branes
are also separated by distances fixed in Planck units.  The light
open string modes stretching between the D0-branes 
have energy $\edkps$, and are therefore relevant in this regime.
Closed strings decouple because
their interactions with the D0-brane quantum mechanics are 
velocity-suppressed.  
Oscillator  modes of open strings also decouple:  their masses
go like $1/l_s= \edkps/g_s^{1/3}$.  
 
In analyzing the compactification, one needs to consider the full spectrum
of states, including wrapped D-branes.  
The hierarchy of masses is as follows:

\eqn\Dsix{{\rm D6-branes:}~~M_6\sim g_s^{2/3} \edkps V_6}
\eqn\open{{\rm stretched ~open ~strings:}~~
	M_1\sim \edkps \delta x}
\eqn\Dfour{{\rm D4-branes:}~~M_4\sim \edkps V_4}
\eqn\osc{{\rm string ~oscillator ~modes:}~~
	M_{osc}\sim{\edkps \over g_s^{1/3}}}
\eqn\Dtwo{{\rm D2-branes:}~~M_2\sim 
	{\edkps \over g_s^{2/3}} V_2}

Here $V_2$,$V_4$, and $V_6$ are the volumes of the two-, four-, and
six-cycles respectively, in eleven-dimensional Planck units (and presumably
a few orders of magnitude at most in these units).
$\delta x$ is the separation between D0-branes, also in eleven-dimensional
Planck units.  As we can see,
in the \lq\lq matrix theory limit", 
one scales out the wrapped D2-branes as well
as the oscillator modes; both have masses which are parametrically larger
than the lowest open string modes as $g_s \to 0$.  

As explained by Sen and Seiberg, the light D6-branes signal a serious problem
in the matrix formulation of this compactification.  They have several
important properties, in addition to their vanishing mass:  

\noindent 1) The BPS multiplet corresponding to the
wrapped D6-brane has states with spin 2 in the transverse dimensions, which 
therefore become
gravitons propagating in the bulk as they go to zero mass.  One
can see this by relating them to D0-branes via T-duality.

\noindent 2) There is a bound state of $k$ D6-branes for any $k$.
Again, T-duality relates this the the bound states of D0-branes.
This identifies the light D6-branes as Kaluza-Klein modes of
a new large dimension.  

Finally, the substringy six-torus with N D0-branes has a T-dual description
in terms of a large six-torus, $\tilde T^6$ 
with N wrapped D6-branes, which we
will denote as $\tilde{D6}$-branes to distinguish them from the wrapped
D6-brane states on the original torus. 
The original D6-branes map to $\tilde{D0}$-branes.
This gives a third important
fact about the matrix formulation of the toroidal compactification:

\noindent 3) It can be described as a 6+1-dimensional theory, at
least before considering the effects of the bulk.
  
This theory does not decouple from the bulk graviton
$\tilde{D0}=D6$ states.  In fact the $\tilde{D6}$-branes,
which start out as Kaluza-Klein monopoles in eleven dimensions,
decompactify in the DKPS limit 
into an ALE space which has no reason to decouple
from the $\tilde{D0}$-brane gravitons \seiberg.

Because the scaling \Dsix\ should hold for any six-manifold
(if we understand how to define $V_p$ and $\delta x$),
it is tempting to conclude that any such background will have
the same problem.  However, as we will now explain,
the matrix compactification
on a generic Calabi-Yau threefold has much simpler properties, and in
particular there is no issue of coupling to Kaluza-Klein gravitons.

\subsec{The DKPS limit of the Calabi-Yau compactification}

We must first determine what the prescription of 
\seiberg\ entails for
the Calabi-Yau case.  We will for concreteness take the Calabi-Yau to
be the quintic hypersurface in $\IP^4$, which has a one-dimensional
K\"ahler moduli space.
In the DKPS limit, we are instructed to study type IIA string
theory on a sub-stringy quintic.  The point in the moduli
space of the quintic where the overall volume shrinks to zero
is the (mirror) conifold point
in K\"ahler moduli space \andyjoe\
(for a nice discussion see e.g. \gk\
and references therein).  Often we will 
slightly abuse terminology 
and refer to the finite distance
singularity in the
K\"ahler moduli space of the quintic as the conifold point. 
We will take the theory of $N$ D0 branes near the conifold
as the definition of matrix theory on the quintic.
This choice will be further motivated in
the following discussion.
In \S2.3\ we will summarize and 
enumerate how points 1),2), and 3) above differ in the
Calabi-Yau case.

In the case of $T^6$, before modding by the $SO(6,6;\IZ)$
modular group one has tori of arbitrarily small radii, which
we can measure via the
masses of string winding modes, and
there is an infinite distance singularity at very small radius.
This is identified by the modular group with a very large
T-dual $\tilde {T^{6}}$.
The analogous discussion for the quintic is more complicated.
If one were to work with the $\it classical$ 
K\"ahler cone,
one would find a very small quintic which is not 
related by any elements of the stringy duality group to
a very large quintic.  However, since the K\"ahler moduli
space receives stringy corrections and the proper
notion of distance is
subtle and probe-dependent \refs{\cydist,\gk},
we should also try to verify our choice for the DKPS limit 
in other ways. 

If we begin by describing the theory as M-theory
on the Calabi-Yau threefold times $S^1$, we would find
that classically, the masses of the wrapped branes and
of the string states scale as in Eqs. \Dsix - \Dtwo.
Thus we would find that the D6-brane is becoming light
and the other branes have finite or large mass relative
to $\edkps$.  
However, the masses will receive corrections
from membrane instantons (in particular, factorization
of the K\"{a}hler and complex structure moduli spaces tells us
that they will be membranes wrapping the $S^1$ and two-cycles of
the Calabi-Yau).  This will generically shift the masses of the
wrapped D2- and D4-branes, and in fact
for the quintic there is no point in the moduli space for
which these masses vanish \gk. 
At the conifold
point the mass of the wrapped D6-brane vanishes even
including such corrections.  

Thus, the hierarchy of branes suggests
that the conifold limit is the correct DKPS/matrix theory limit.  Still,
one could ask why this is true intrinsically.  
This is the only point at which any size is truly going to zero,
rather than being of order $\ell_s$.\foot{In particular, if
one chose any other point in the K\"ahler moduli space,
the situation would only $\it improve$ -- there are no extra light
degrees of freedom.  In this sense, we are making the most 
dangerous choice and arguing that the matrix theory is still
simple.}  
But this direct discussion is subtle due to the worldsheet instantons.

We can in fact avoid complications coming from worldsheet instantons,
by going to
the mirror quintic to discuss the limit.  On the mirror, we work
with the complex structure moduli space, which
receives no worldsheet instanton corrections. 
There, before dividing by the modular
group of the complex structure moduli space,  
there is only one infinite distance
\lq\lq large complex structure" limit.  The
conifold locus is not identified with
any other infinite distance singularity, as it is at
finite distance \cdgp.  The conifold locus is the 
only special point mirror to a \lq\lq small" Calabi-Yau, and 
so we expect it to be 
singled out as the DKPS/matrix theory limit.
This point of view again supports our conclusion that
the DKPS limit on the quintic involves a 
quintic near the conifold point, and in addition it
cannot be
\lq\lq T-dualized" to a large quintic. 

Of course, one $\it can$ use mirror symmetry to relate the
theory of N D0 branes on the original quintic to N $\tilde{D3}$ 
branes 
wrapping the fibers of a $\tilde{T^3}$ fibration on the mirror
quintic \syz.  This gives a 3+1 dimensional theory,
instead of the 6+1 dimensional theory which arises in 
$T^6$ compactification.

In the limit we have defined, we have only
the D6-branes becoming light, and this gives rise to precisely
{\it one} massless hypermultiplet, as in \andy.  
There are no multi-D6 brane bound states, as
we can see by mirror symmetry; the mirror of
the wrapped D6-brane is the D3-brane wrapped around
the vanishing three-cycle \refs{\andyjoe, \gk}.
Consistency requires that there only be one such state \andy\
and this was indeed found to be the case \bsv.
Also, one finds
$\it no$ extra massless particles from wrapped D2 or D4 branes
in this limit.  Although the overall volume has shrunk to
zero, the volumes of 2 and 4 cycles in this limit are actually of 
order $l_s^{2}$ and $l_{s}^{4}$, respectively \gk.  
This surprising assertion is a property of
the quantum corrected volume in compactified string theory, 
and is a direct consequence of mirror symmetry. 

We should also ask about the masses of the light open string modes.
If we start at large radius with
N D0 branes separated by sub-stringy distances, and then shrink
the Calabi-Yau, we might expect the lightest stretched strings
between the D0 branes to remain much lighter than 
the oscillator modes. This is because we do not expect
that the effective distance between the zero branes would
{\it increase} as we shrink the Calabi-Yau.  (For
example, the sizes of two-cycles are bounded below for the quintic,
but they do decrease monotonically as we move from
the large-radius limit to the conifold point).  Still, at
the (mirror of the) conifold point, worldsheet instanton
effects are strong and to understand what degrees of freedom
are present, one will have to understand the effects of string 
instantons.  It is perhaps easier to think about this problem
in the 3+1 dimensional theory on the mirror.

On the mirror, the D0 branes map to $N$ $\tilde{D3}$ branes wrapping
the fiber $\tilde{T^3}$, while the base $S^3$ is becoming very small --
the moduli are the locations of the $\tilde{T^3}$ on the base and the 
Wilson lines of the D-brane theory around the cycles of $\tilde{T^3}$
\syz.   
We can be at a generic point in the K\"ahler moduli space of the
mirror,
so in particular (for \lq\lq large enough" complex 
structures on the quintic)
we can take worldsheet instanton effects to be
suppressed on the mirror.  Then, 
distance is measured using the classical
metric.
In this set-up, the stretched strings between three-branes which
are very close together on the base $S^3$ are much lighter than
the string oscillator modes, because their masses are well approximated
by the naive classical formula.  These lightest strings
give the $W$ bosons of the 3+1 dimensional quantum field theory,
and although their mass might be renormalized by 
a significant numerical factor
it will not be parametrically enhanced to $M_s$. 

\subsec{Problems (1)-(3) revisited}

We have argued in the previous subsection that the conifold regime
is the DKPS regime we should use to define the DLCQ quantization of
matrix theory on the Calabi-Yau.  Given this, there are very precise
differences between the $T^6$ case and the Calabi-Yau case.
In particular,
the properties 1)-3) of $T^6$ compactifications
which complicate their
matrix theory description have very different analogs
in the Calabi-Yau case.

\noindent 1) On the Calabi-Yau, 
the D6-brane has the quantum numbers of a {\it hypermultiplet}, instead
of a gravity multiplet,
in the transverse spacetime.  Indeed, the light D6-brane
is the monopole
state that resolves the conifold singularity in the IIA string theory
compactification on the CY \andy.  This also implies: 

\noindent 2) Here the wrapped D6-branes do not form multi-sixbrane bound
states, unlike in the torus case.

\noindent 3) Here there is no T-duality symmetry (more precisely, no
element of the discrete symmetry group of the CY compactification) which
maps D0-branes on the CY in the conifold regime to $\tilde{D6}$-branes
on a large CY.  
The conifold singularity is at
finite distance from the interior points on the moduli space \cdgp,
whereas the large radius singularity is at infinite distance.
Therefore there is no reason to believe the physics is
that of a 6+1-dimensional theory.\foot{
For Calabi-Yau compactifications 
realized as orbifolds of $T^6$,
one may perform T-duality on all six directions of the orbifold, so
as noted above the story will differ from the cases we are discussing
here.}
One $\it can$ 
mirror symmetrize so that the D0-branes become 
$\tilde{D3}$-branes wrapping
a $\tilde{T^3}$ on the mirror Calabi-Yau \syz.  
This leaves us with a 3+1-dimensional theory, which can 
be well defined without gravity and which by
the arguments we have given does
not seem to couple to gravity.

We see from 1) and 2) that the matrix description of Calabi-Yau 
compactifications does not involve gravity, and the limit of 
\refs{\sen,\seiberg} 
does not entail the growth of an extra dimension.  We will present tentative
arguments that the light D6-brane may also decouple from the D0-branes
in the next section.  Note that even if this is not the case, the matrix
formulation is a considerable simplification over the original description
of the M-theory compactification, as the relevant degrees of freedom
do not seem to include gravity.

\newsec{Coupling of the D0 branes to the Wrapped D6 Hypermultiplets}

Although the matrix description does involve a considerable
simplification, we still should wonder to what extent the
heavy D0 branes will couple to the light wrapped D6 branes.
This issue is still not clear to us.  However, we will present
two indications that the D6-branes might in fact decouple from the
D0-brane theory in this case.

One indication is that the open string theory that lives on
the D0-branes (perhaps mirror symmetrized to $\tilde{D3}$-branes)
may be consistent by itself.  
If this theory is non-singular, then we would
not expect the theory to contain
non-perturbative light states. 
Before considering this open string theory in
the DKPS limit,
let us recall first how things worked for the ordinary
IIA string theory compactified
on the Calabi-Yau.  This closed string
theory becomes ill-defined at the conifold singularity \cdgp.  
One can see this using mirror symmetry as in \cdgp.  Another very
useful way to get a handle on such singularities is to study the
compactification using the gauged 
linear sigma model technique introduced
in \edphases.  There, the signal of the singularity is the emergence
of a noncompact branch (a throat) in the target space of the two-dimensional
sigma model (indeed, as argued in \edphases, the appearance
of a noncompact direction is the only possible
source of a singularity in such models).  
As in \edphases\ we will refer to the complex coordinate on this
branch as $\sigma$.  
The field $\sigma$ is a neutral scalar in the $U(1)$ gauge multiplet
of two-dimensional $(2,2)$ supersymmetry.
The existence of the $\sigma$ branch does not by
itself imply that the closed-string physics is singular, but this
turns out to be the case.  In particular, one finds that $\sigma$ is
the vertex
operator for the closed string state corresponding to the K\"ahler modulus, 
whose correlation functions
diverge at the singularity \silvwitt.  In general, there are multiple fields
$\sigma_a,~a=1,\dots,h^{1,1}$ corresponding to the elements of $H^{1,1}$.  
The wavefunctions for other
states (for example complex structure moduli) are not supported down
the various $\sigma$ branches.

Now let us consider the situation for the open strings.  This
can be approached by studying the linear sigma model of \edphases\ on
worldsheets with boundaries \albionetal.  In principle we need to
undertsand the details of this model to analyze the singularity structure,
but we can already make a strong qualitative argument.
One finds that at the singularity
the open string theory also has an extra branch, where the open strings 
have Neumann boundary conditions \albionetal.  Here it is
one real dimensional; we will call it $\sigma_R$, as it arises from the
real part of $\sigma$.  But in this case the open string
states do not correspond to elements of $H^{1,1}$, and therefore have
no natural relation to the $\sigma_a$ fields.  In particular, at large
radius the vertex operators for the open strings are linear in
the charged fermions $\psi$ (or their bosonic partners $\partial\phi$
depending on the picture).  Since they are charged
these vertex operators are
never just
proportional to the neutral field $\sigma_a$, even at small radius.    
The corresponding wavefunctions of open string states
are suppressed down the $\sigma_R$ branches since the
charged fields become extremely massive for large $\sigma$;
thus, we do not expect these vertex operators 
to have singular correlation functions at the conifold singularity.
This suggests that the open-string theory living on the D0-branes is
consistent by itself, without extra light degrees of freedom such
as the D6-branes which propagate in bulk.

Another way to approach this issue is to study the D0- and D6-brane
states in the low-energy effective action of the spacetime theory.
D0- and D6-branes
carry dual electric-magnetic charges in the low-energy
N=2 supersymmetric Yang-Mills theory.
From this point of view, an analogous question 
has been studied in quantum field theory.
We can think of
the heavy D0 branes as being very massive magnetic monopoles,
and the light D6 branes as being light electrically charged
particles.  The question is then, to what extent do the heavy
and slowly moving monopoles create electron-positron pairs?
Although the analogy is not precise (the D0 branes have spin 2, and
move around on a Calabi-Yau target space!), 
we can get some intuition by considering this case.
The suppression of the Schwinger pair production 
effect for small fields (and
therefore small monopole velocities) suggests that the bulk D6-branes
are not pair-produced by the D0-branes in our setup.

\newsec{Conclusions}

We conclude that in the DKPS limit of Calabi-Yau compactifications, 
which we are instructed to
take in order to find a matrix formulation \refs{\sen,\seiberg},
there is no sign of the infinite tower of light gravitons which
complicate the matrix description of $T^6$ compactification.
Instead, there is a single massless hypermultiplet \andy.  
So although
the matrix description is by no means trivial, it is a considerable 
simplification over the spacetime theory which contains gravity. 

At this point we should begin to ask what sort of theory
this limit describes.  Now that we know that states
which would make the theory intractable do not
haunt us, we need to know what states to include.
One might ask, for example, whether the DLCQ of
this theory has a finite or infinite number of degrees of freedom --
in other words, is it a higher-dimensional field theory or 
non-gravitational
string theory, and if so
in how many dimensions does it live?

In the case of the torus, we know that we have 6+1-dimensional theory
because there is an infinite tower of winding
modes corresponding to each dimension of the torus,
with masses which are multiples of $\edkps$.  These
can then be described as momentum modes
in some dual theory.  

For curved backgrounds we do not have topologically stable
winding sectors, but we may still have stationary
string wavefunctions corresponding to strings with non-trivial
spatial extent and enough of these could give us a field theory's
worth of degrees of freedom.  As an example, there seems
to be a field-theoretic description of compactification
on a general (i.e. non-orbifold) K3 surface which one may
derive via a Fourier-Mukai transformation corresponding
to T-duality in all four directions of the K3 \horioz.  The
field theory should have low-energy modes
corresponding to excitations of the gauge theory, and
they should be dual to states of stretched open strings
(since T-duality generally exchanges kinetic and tensile
energy of the string).

It is not completely clear what happens in our case, as the string theory
lives in the regime where quantum geometry is very important.
There is a significant difference between our case and the
torus and K3 cases.  For us the DKPS limit (the conifold point on
the moduli space) is at finite distance.  On the torus and on K3,
the relevant singularities are at infinite distance.  This is related
to the tower of winding strings coming down to zero mass there.

Ideally, one would be able to look at the annulus
diagram for the open string states and count the states which
have energies of order $\edkps$ or less;
we could do this
by measuring the conformal dimensions of
the Virasoro primaries as a function of the K\"ahler
moduli, as we approach the conifold.  Of course, in the quintic
the only regimes where we know how to do this
calculation are at the large radius limit and at the Landau-Ginzburg
orbifold point.  Indeed, some preliminary work has been done
on describing D0-branes in the latter case \gepneropen.
Since we are interested in the conifold regime, these calculations
do not directly apply to our setup.

Another issue which deserves study is the re-emergence
of classical geometry.  Our goal is
the description of M-theory, rather than string theory, compactified
on a Calabi-Yau.  At finite $N$ we describe the theory
via type IIA with vanishing string coupling, and it appears that
membrane instantons wrapped around $S^1$ times a two-cycle
are important.  In the large $N$ limit such effects should go away,
and we will need to understand the mechanism by which such effects
decouple.

Finally, we should note that we have not addressed the
problems pointed out in \refs{\dos, \do} with compactifications
that break supersymmetry.  
It seems that one sensible interpretation, following \hellpol, is
that there is no reason to expect a simple correspondence between
supergravity and matrix theory computations for finite $N$.

It is tempting to speculate that the simplification with respect to
the $T^6$ case means that the
matrix model in some sense
prefers backgrounds with reduced supersymmetry. 
It will be interesting to consider further supersymmetry breaking
down to $4d$ N=1 supersymmetry.  There instanton effects
will be of interest, but one can again begin
in the DKPS regime in type I' theory on a Calabi-Yau.  In particular,
worldsheet instantons of the type I/type I' theory do not
contribute to the superpotential and therefore do not lift
radial moduli.
Of course there are
also qualitatively different problems in light-cone gauge  
compactifications down to four and fewer dimensions. 
Perhaps the story there will also depend in a crucial way on the
number of unbroken supercharges in a given compactification.

\bigskip
\centerline{\bf Acknowledgements}

We would like to thank O. Aharony, V. Balasubramanian,
T. Banks, R. Gopukamar, J. Maldacena, D. Morrison, 
A. Rajaraman, S. Shenker, and A. Strominger 
for useful discussions and comments.  E.S. would like to
thank the physics department at Harvard for hospitality during
part of this project.  
The work of S.K. is supported by NSF grant PHY-95-14797, 
by DOE contract DE-AC03-76SF00098, and by a DOE Outstanding
Junior Investigator Award. The work of A.L. is
supported by NSF grant PHY-92-18167.
The work of E. S. is supported by the
DOE under contract DE-AC03-76SF00515.

\listrefs
\end